\documentclass[aps,nobibnotes,nofootinbib,prl,reprint,groupedaddress]{revtex4-1}
\def\beq{\begin{eqnarray}}
\def\eeq{\end{eqnarray}}

\def\ba{\begin{eqnarray}}
\def\ea{\end{eqnarray}}
\def\beq{\begin{eqnarray}}
\def\eeq{\end{eqnarray}}

\def\mpl{M_{\rm Pl}}

\def\p{{\cal P}}

\def\L*{{\cal L}_*}
\def\L{\mathcal{L}}
\def\({\left(}
\def\){\right)}

\def\nn{\nonumber}
\def\p{\partial}
\def\mn{_{\mu \nu}}

\def\p{\partial}

\def\<{\langle}
\def\>{\rangle}

\usepackage{hyperref}
\usepackage{amsmath}
\usepackage{amsfonts}
\usepackage{verbatim}
\usepackage{setspace}
\usepackage{url}
\usepackage{color}
\usepackage{subfig}
\usepackage{graphics}
\usepackage{graphicx}
\def\be{\begin{eqnarray}}
\def\ee{\end{eqnarray}}
\def\beq{\begin{eqnarray}}
\def\eeq{\end{eqnarray}}

\def\beq{\begin{eqnarray}}
\def\eeq{\end{eqnarray}}

\def\mpl{M_{\rm Pl}}

\def\p{{\cal P}}

\def\L*{{\cal L}_*}
\def\L{\mathcal{L}}
\def\({\left(}
\def\){\right)}

\def\nn{\nonumber}
\def\p{\partial}
\def\mn{_{\mu \nu}}

\def\p{\partial}

\def\<{\langle}
\def\>{\rangle}

\def\be{\beta}
\def\lsim{\mathrel{\rlap{\lower3pt\hbox{\hskip0pt$\sim$}}
     \raise1pt\hbox{$<$}}}         
\def\gsim{\mathrel{\rlap{\lower4pt\hbox{\hskip1pt$\sim$}}
     \raise1pt\hbox{$>$}}}         
\def\lsim{\mathrel{\rlap{\lower3pt\hbox{\hskip0pt$\sim$}}
     \raise1pt\hbox{$<$}}}         
\def\gsim{\mathrel{\rlap{\lower4pt\hbox{\hskip1pt$\sim$}}
     \raise1pt\hbox{$>$}}}         

\def\x{\textbf{x} }

\begin{document}


\title{}


\title{Selfacceleration  with  Quasidilaton}
\author{Gregory Gabadadze$\,^*$, Rampei Kimura$\,^\dagger$, David Pirtskhalava$\,^\ddagger$}
\affiliation{$^*$Center for Cosmology and Particle Physics,
Department of Physics, New York University, New York,
NY, 10003, USA \\
$^\dagger$Research Center for the Early Universe,
The University of Tokyo, Tokyo 113-0033, Japan \\
$^\ddagger$Scuola Normale Superiore, Piazza dei Cavalieri 7, 56126, Pisa, Italy
}



\begin{abstract}

Quasidilaton massive gravity is an extension of massive General Relativity to a theory with
additional scale invariance and approximate internal Galilean symmetry.
The theory has a novel self-accelerated solution with the  metric indistinguishable (in the decoupling limit) from 
the de Sitter space, its curvature set by the graviton mass.  The spectra of tensor, vector,  and scalar perturbations on this solution  
contain neither ghosts, nor gradient instabilities  or superluminal modes,  for a range of 
the parameter space.  This represents an example  of a self-accelerated solution  
with viable perturbations, attainable within a low energy effective field theory.

 \end{abstract}


\maketitle

\paragraph{ \it \bf Introduction and Summary:}

Both the existence and the magnitude of late-time cosmic acceleration could be hinting to  a 
modification of gravity at cosmological distances. Of such modifications, massive gravity is arguably 
one of the best motivated.  While the Fierz-Pauli (FP) action \cite{Fierz:1939ix} gives the unique  theory of 
a free spin-2 state of mass $m>0$, generalization to a nonlinear theory  has proven to be hard 
due to the emergence of the Boulware-Deser (BD) ghost  \cite{Boulware:1973my}. 
A special set of the  graviton mass and  nonlinear potential terms  
proposed  in \cite{deRham:2010ik,deRham:2010kj} (referred to as the dRGT terms)  eliminates the BD ghost,  
as shown  perturbatively to the quartic order in \cite{deRham:2010kj}, and  to all 
orders in the unitary gauge \cite{Hassan:2011hr,*Hassan:2011ea}, and beyond 
\cite {Mirbabayi:2011aa,*Hinterbichler:2012cn, *Deffayet:2012nr,*Kugo:2014hja}.  
The resulting quantum theory can be viewed as an effective field theory with 
the (Minkowski-space) strong coupling scale given by 
$\Lambda_3=(\mpl m^2)^{1/3}$ \cite{deRham:2012ew,*deRham:2013qqa}.

Massive gravity possesses the self-accelerated (de Sitter) vacua with curvature $\sim m^2$
 \cite{deRham:2010tw,Koyama:2011xz,*Nieuwenhuizen:2011sq,D'Amico:2011jj}.  
Unfortunately, kinetic terms of either scalar or vector perturbations vanish  on the selfaccelerated  
backgrounds \cite{deRham:2010tw,Khosravi:2013axa,*DeFelice:2013bxa}\footnote{Despite of there being no reason why these kinetic terms would not be generated quantum mechanically - \emph{e.g.} via matter loops - their vanishing at the classical level could be discouraging.}.
A way out might be to promote the graviton mass to a dynamical field that gives rise to a new 
nonlinearly realized global symmetry, as well as an approximate internal Galilean invariance \cite{D'Amico:2012zv}. 
This extension, dubbed quasidilaton massive gravity (QMG), retains 
freedom from the BD ghost.  A special class of selfaccelerated solutions of  QMG  
found  in \cite{D'Amico:2012zv},  suffer from a 
major problem:  one of the perturbations  of the theory  necessarily flips the sign of its kinetic 
term on these backgrounds \cite{Gumrukcuoglu:2013nza,D'Amico:2013kya}. 
In addition, the scale of time variation of the 
quasidilaton field on these special solutions is  much higher than the strong coupling 
scale, $\sqrt{\mpl m} \gg \Lambda_3$   \cite{D'Amico:2013kya}, making 
them vulnerable to unknown high-energy physics  (i.e.,  the special solutions cannot be 
obtained  in the decoupling limit \cite {D'Amico:2012zv,D'Amico:2013kya}).  QMG 
has been recently extended by De Felice and Mukohyama  \cite{DeFelice:2013tsa,*DeFelice:2013dua} by  derivative operators 
consistent with the scale symmetry;  these  have an effect of flipping the sign of the wrong 
kinetic term back to normal, hence solving the major problem.  While certainly an important proof of a concept, 
the De Felice-Mukohyama approach does still leave  the issue of UV sensitivity open  \cite {D'Amico:2013kya}.   
The purpose of this letter is to present novel selfaccelerated solutions in QMG with viable perturbations, all 
attainable within the low energy field theory.

{ \it \bf The theory and its decoupling limit:} We work in the decoupling limit (DL), 
defined as $ \mpl\to \infty,~ m\to 0,~ \Lambda_3=\text{fixed}$.
The  theory then captures physics at distances ranging from $\Lambda^{-1}_3\sim~ 1000 $ km, all the way up to 
(almost) $m^{-1}\sim H^{-1}_0\sim 10^{23}\, {\rm km}$ (see, e.g. \cite {deRham:2010tw,D'Amico:2011jj})\footnote{Under certain meaningful assumptions about the UV completion, the distance scale $\Lambda_3^{-1}$ can be significantly reduced for realistic astrophysical backgrounds \cite{Nicolis:2008in,Nicolis:2004qq,*Nicolis:2009qm,Berezhiani:2013dca}.}.

The relevant dynamical degrees of freedom that enter into the QMG action in the DL are the helicity-2 ($h_{\mn}$), helicity-1 ($A_\mu$) and helicity-0 ($\pi$) states of the massive graviton, as well as the quasidilaton field $\sigma$. 
With the recent developments, a closed-form expression for the vector-scalar interactions can be obtained by using  a non-dynamical antisymmetric tensor $B_{\mn}$ \cite{Gabadadze:2013ria,Ondo:2013wka}.  The matrix of the helicity-0 second derivatives is denoted by $\Pi_{\mn}=\p_\mu\p_\nu\pi$. The ordered index contractions on the epsilon symbols  
will be implied\footnote{ We use the mostly plus signature and all indices are manipulated by  
the flat metric $\eta_{\mu\nu}$  and its inverse.  The Levi-Civita symbol,
$\varepsilon_{\mu\nu\alpha\beta}$,  is normalized so that  $\varepsilon_{0123}=1$.}; for example, 
$\varepsilon_{\mu_1\mu_2\mu_3\mu_4}\varepsilon^{\nu_1\nu_2\mu_3\mu_4}
\Pi^{\mu_1}_{~\nu_1}\Pi^{\mu_2}_{~\nu_2}
\equiv\varepsilon\varepsilon\Pi\Pi$, as well as 
$\varepsilon_{\mu\mu_2\mu_3\mu_4}\varepsilon^{\nu\nu_2\nu_3\mu_4}
\Pi^{\mu_2}_{~\nu_2}\Pi^{\mu_3}_{~\nu_3}\equiv\varepsilon_\mu\varepsilon^\nu
\Pi\Pi$, with an obvious generalization to terms with a different number of $\Pi$'s. The placement of indices should be treated with care for expressions involving the auxiliary tensor $B^\mu_{~\nu}$, its square $(B^2)^\mu_{~\nu}=B^\mu_{~\alpha}B^\alpha_{~\nu}$, and (the first derivative of) the vector helicity $\p_\mu A^\nu$. For example, $\varepsilon\varepsilon B\p A$  denotes  $\varepsilon_{\mu_1\mu_2\mu_3\mu_4}\varepsilon^{\nu_1\nu_2\mu_3\mu_4}
B^{\mu_1}_{~\nu_1} \p_{\nu_2} A^{\mu_2}$.   With these conventions, and setting $\Lambda_3=1$, the Lagrangian describing the decoupling limit of QMG without a potential can be obtained and is  written as follows:
\begin{widetext}
\ba
\label{eq:mgmasteraction}
\mathcal{L}_{DL} &=&
-\frac{1}{2} h^{\mn} \( \mathcal{\hat E} h \)_{\mn}
+h^{\mu\nu} \left[- \frac{1}{2} \varepsilon_\mu \varepsilon_\nu \Pi
+  a_2\varepsilon_\mu \varepsilon_\nu \Pi \Pi
+  a_3\varepsilon_\mu \varepsilon_\nu \Pi \Pi \Pi \right] +\frac{1}{\mpl}h^{\mn} T_{\mn}
-\frac{1}{4}\bigg[2 \varepsilon\varepsilon BB +(4-8 a_2) \varepsilon\varepsilon BB\Pi \nn \\ &-&4(a_2+3a_3) \varepsilon\varepsilon BB\Pi\Pi+2\varepsilon\varepsilon B^2\Pi -4a_2\varepsilon\varepsilon B^2\Pi\Pi-4a_3 \varepsilon\varepsilon B^2\Pi\Pi\Pi+4\varepsilon\varepsilon B\p A 
-16 a_2\varepsilon\varepsilon B\p A\Pi -24 a_3\varepsilon\varepsilon B\p A\Pi \Pi\bigg]\nn \\
&-&\omega \p^\mu\sigma\p_\mu\sigma+\sigma\bigg[ \varepsilon\varepsilon\Pi-2 (a_2+1)\varepsilon\varepsilon\Pi\Pi+\frac{2}{3}(4a_2-3a_3+2)\varepsilon\varepsilon\Pi\Pi\Pi-\frac{1}{3}(2a_2-6a_3+1)\varepsilon\varepsilon\Pi\Pi\Pi\Pi      \bigg]~.
\ea
\end{widetext}
Here $\mathcal{\hat E}$  is the Einstein operator, and $a_{2,3}$ and $\omega$ are free parameters of the theory.  The $\pi$-$h$ interactions are those of pure massive gravity \cite{deRham:2010ik}, while the $\sigma-\pi$ interactions are (bi-) Galieons \cite{Nicolis:2008in}. The action is invariant under $U(1)$ gauge transformations, $A_\mu\to A_\mu+\p_\mu\alpha$. The external stress-tensor couples to the helicity-2 field $h_{\mn}$, and the latter determines geodesics of ordinary matter.  We will be interested in self-accelerated backgrounds ($T_{\mu\nu}=0$) with the Hubble constant $H$;  for these  $\bar h_{\mn}=-(H^2 x^2/2)\eta_{\mn}$ and  
$\bar \pi=qx^2/2 \quad (q=\text{const})$, while the solution  for 
$\sigma$  can in general take the form  $\bar\sigma=(-q_0t^2+q_1 \textbf{x}^2 ) /2$. 
We will focus on the solution with $\bar B^\mu_{~\nu}=0,~\bar A_\mu=0$. 
Then, the equations for the helicity-2,  quasidilaton,  and the helicity-0 fields give respectively
\begin{widetext}
\begin{align}
\label{eq:eqfr}
 H^2-2q\(-\frac{1}{2•} + a_2 q+a_3 q^2\)&=0~,\\
\label{eq:eqsigma}
(2 a_2-6a_3+1) q^4-2 (4a_2-3a_3+2) q^3 + 6(a_2+1)q^2-3q+\frac{\omega}{4•}q_t&=0~,\\
\label{eq:eqpi}
 H^2\(3 a_3 q^2+2 a_2q-\frac{1}{2}\) +q_t \bigg[q^3\(\frac{2}{3•}a_2 -2 a_3+\frac{1}{3•}\) - q^2 \(2 a_2-\frac{3}{2} a_3+1\)
+q (a_2+1)-\frac{1}{4•}  \bigg]&=0~,
\end{align}
\end{widetext}
where $q_t\equiv q_0+3q_1$, and  $q_0$ and $q_1$ enter only via $q_t$, meaning that $\sigma$ can 
be shifted by  a zero mode of the d'Alambertian, 
$\sigma\to\sigma-\alpha t^2+\beta \x^2$,  $\alpha+3\beta=0$,   without changing the background.
Below  we will work with $q_t$ and $q_1$; the former is determined from the equations, while $q_1$ is a 
free 'flat direction'. Nevertheless,  $q_1$ will affect  perturbations, considerably expanding the acceptable parameter space.  The equations are all linear in $q_t$ and $H^2$, therefore these can be expressed in terms of $q$ via \eqref{eq:eqpi} and \eqref{eq:eqfr} to obtain an algebraic master equation for $q$  from \eqref{eq:eqsigma}. The resulting equation is of the seventh order and in general not soluble analytically.

\paragraph{ \it \bf Perturbations:} We start with the vector perturbations of  \eqref{eq:mgmasteraction}. Recalling that $\p_\mu\p_\nu\bar\pi=q\eta_{\mn}$  and expanding the epsilon symbols, one finds $$\mathcal{L}_V=Q_1 B_{\mn} B^{\mn}+Q_2 B_{\mn} F^{\mn}~,$$ where $Q_2=Q_1/(q-1)=6a_3q^2+4 a_2q-1$.  In the absence of the quasidilaton, the $\pi$-equation of motion obtained from \eqref{eq:eqpi} by setting $q_t=0$, implies that $Q_{1,2}=0$, leading to infinitely strong coupling of the vector perturbations \cite{deRham:2010tw}. In QMG however, $\pi$-equation is modified by extra terms  and  
no longer leads to this issue: one gets dynamical and weakly coupled (in the infrared) perturbations, propagating at the speed of light.  Integrating out the $B_{\mn}$ field, the following necessary and sufficient condition for the absence of vector ghosts is obtained
\beq
\label{eq:vecghost}
\label{eq:cond2}
Q_1=(q-1) (6a_3q^2+4 a_2q-1)>0~.
\eeq
The scalar perturbations are described by  the following  Lagrangian
\begin{align}
\label{eq:scalarlag}
\mathcal{L}_s=A_1 \delta\dot \pi^2 -A_2 (\p_i\delta\pi)^2&+B_1 \delta\dot\pi\delta\dot\sigma-B_2 \p_i\delta\pi\p_i\delta\sigma\nn \\
&+C_1 \delta\dot \sigma^2-C_2 (\p_i\delta\sigma)^2~,
\end{align}
where the coefficients $\(A,B,C\)$ are given as follows:
\ba
\label{eq:pert}
A_1=6 H^2 (3 a_3 q+a_2)+6 c^2 ~~~~~~~~~~~~~~~~~~
 \nonumber \\ + 12 q_1\bigg[ (2 a_2- 6 a_3+1)q^2 + (3 a_3-4 a_2-2)q+a_2+1   \bigg],   \nonumber \\
A_2=6 H^2 (3 a_3 q+a_2)+6 c^2 
+ 4(q_t-q_1) ~~~~~~~~~~\nonumber \\ \times \bigg[ (2 a_2-6 a_3+1)q^2 + (3 a_3-4 a_2-2)q+a_2+1   \bigg], \nonumber \\
B_{1}=B_2= 8 (2 a_2-6a_3+1) q^3 - 12 (4 a_2-3 a_3+2)q^2~~~  \nonumber \\ +24 (a_2+1) q
-6 ~, \qquad
C_1=C_2=\omega~.~~~~~~~~~~~~
\ea
For a Lorentz-invariant quasidilaton profile with $q_1=q_0$, one gets $A_1=A_2$  and both scalar  
modes propage at the speed of light. Relaxing this condition, as we will see shortly, results in a 
deviation from unity of the speed of sound for only one of the two modes.

\paragraph{ \it \bf Stability and (sub)luminality:}
As noted above, it is impossible to obtain \emph{the most general} closed-form analytic expressions for the viable parameter space. One can however employ certain limits and/or numerical study to show the existence of such. To this end, it is simpler to trade the free parameter $a_2$ for $q$, since $a_2$ can be algebraically expressed in terms of $q, a_3, q_t$ and $\omega$; for any $q$ then, one can straightforwardly find a theory with the corresponding value of $a_2$. 
We will first consider the Lorentz-preserving case, $q_1=q_0=q_t/4$. Expressing $H^2$ (via the Friedmann equation) and $a_2$ (via the $\sigma$ equation) in terms of the rest of the parameters, one finds a \emph{quadratic} equation for 
$q_t$, that depends on $q, a_3$ and $\omega$ -- the only remaining free parameters of the theory. 
The necessary condition for the absence of ghosts in the scalar sector is $A_1>0, ~~C_1>B_1^2/(4 A_1)$. Let us concentrate on the limit of \emph{large} $q$. Then, one can straightforwardly find the quadratic equation from which $q_t$ can be determined,
$$-4\omega q^4 q_t^2+72 a_3 q^7 q_t+288 a_3^2 q^{10}=0~.$$
The two solutions are\footnote{We have assumed $a_3 q^3>0$, since this has to be the case for a positive $H^2$, as shown shortly.}
$ q_t={3 a_3 q^3 \(3\pm \sqrt{8\omega+9} \)} /{\omega}.$
The second of these (featuring the 'minus' sign) does not lead to any acceptable parameter space, so we concentrate on the first one. The value of $a_2$ for this solution is\footnote{One has to retain a subleading piece for $a_2$ due to a cancellation of leading terms in the the expressions for perturbation coefficients $A_1$ and $B_1$ in \eqref{eq:pert}.}
\beq
a_2=3 a_3-\frac{1}{2} +\frac{3a_3 \(21 -\sqrt{8\omega+9}  \)}{8 q}+\mathcal{O}\(\frac{1}{q^2}\),\nn
\label{a23}
\eeq
while the (leading-order) expressions for the Hubble parameter squared and the normalization coefficient of the vector gauge kinetic term, $Q_1$, are $H^2=2 a_3 q^3$, and $Q_1=6 a_3 q^3$. Both are positive for $a_3 q^3>0$, leading to dS backgrounds with ghost-free vector perturbations.  With the above expressions used, the quantities determining stability of the scalar perturbations are then given by
\ba
A_1&=&\frac{9 a_3^2 q^4 \(8\omega+27+9\sqrt{8\omega+9}\)}{2\omega},\nn \\
C_1-\frac{B_1^2}{4 A_1}&=& \omega-\frac{2\omega (\sqrt{8\omega+9}-3)^2}{8\omega+27+9\sqrt{8\omega+9}}~.
\ea
Both are positive for $0<\omega<54$. The parameter space (for $q\gg1$) corresponding to ghost-free self-accelerated backgrounds is therefore given by 
\beq
\label{a24}
0<\omega<54, \qquad \operatorname{sgn}(a_3)=\operatorname{sgn}(q)~.
\eeq 
All modes propagate at the speed of light, excluding any gradient instabilities\footnote{Both 
the $\omega \to 0$  and $\omega \to 54$ limits give  an infinitely strongly coupled theory. To avoid this, one can use, say, $1/2< \omega <53$~.}. 
This provides an analytic proof of the existence of fully 
stable dS backgrounds for  large $q$.   The latter here is parametrically greater than unity (say, $q\sim 10$)  due to the  arrangement between  $a_{2,3}$,  while the solution itself 
belongs to the low energy field theory,  as it is obtained in the decoupling limit.     
\begin{figure}
\includegraphics[width=.22\textwidth]{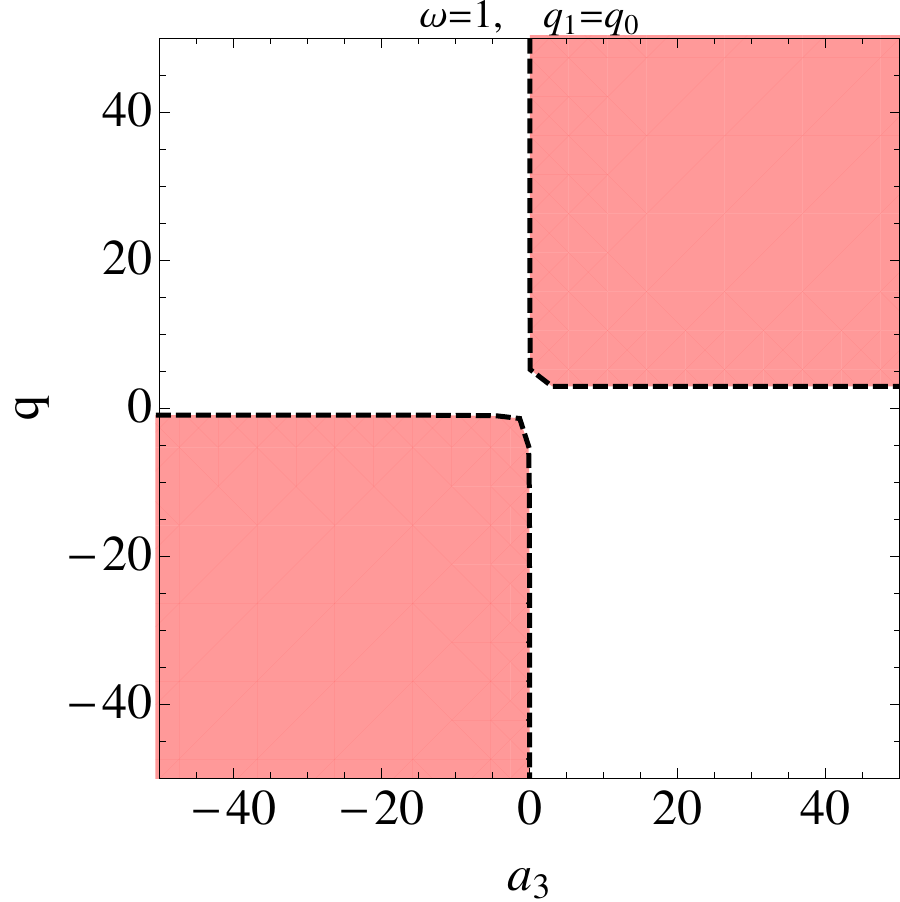} \quad
\includegraphics[width=.22\textwidth]{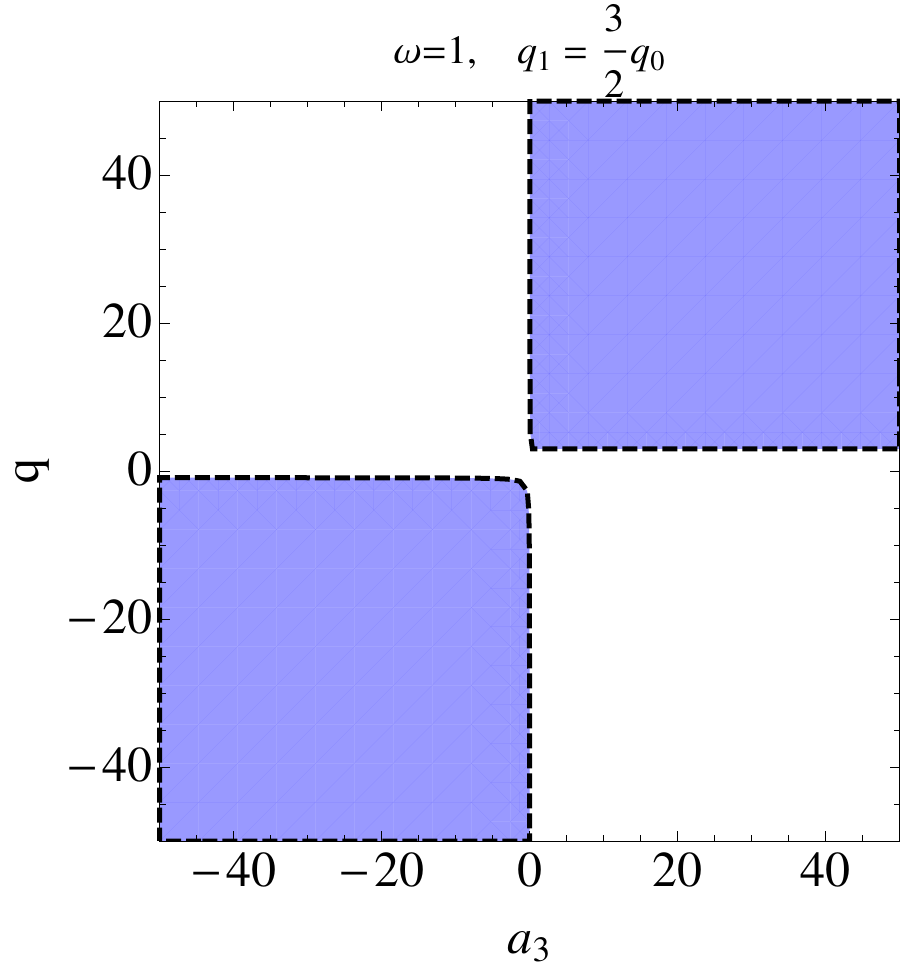}
\caption{Left: a subset of the parameter space for stable de Sitter solutions with a Lorentz-invariant $\sigma$ profile (both scalar modes are exactly luminal). Right: the parameter space supporting stable dS backgrounds with $q_1=3 q_0/2$. One of the scalars is strictly subluminal.}
\label{fig:2}
\end{figure}
A numerical analysis shows the existence of stable self-acceleerated vacua for intermediate values of $q$ as well. One representative subset of the allowed parameter space is given in the left panel of Fig.\ref{fig:2} (note that the allowed space does \emph{not}  extend to $|q|\sim 0$; 
instabilities appear for $|q|\ll 1$). On these  plots we used  $\omega=1$. The colored region corresponds 
to one corner of the full parameter space for which the conditions of absence  of the ghosts, 
gradient instabilities and superluminal propagation are  all satisfied. 

We have also explicitly checked the case  $q_1\neq q_0$; the stable de Sitter vacua are retained:
the two speeds of sound  of the scalar modes  in this case are $c^2_s=\(x\pm\sqrt{x^2-4 y z}\)/2 y ~,$
where $x=4 A_1C_2+4 A_2C_1-2 B_1B_2,~ y=4 A_1C_1-B_1^2,$ and $z=4 A_2C_2-B_2^2~$. 
Using the expressions in \eqref{eq:pert}, we conclude that one of the scalar 
modes always propagates at the speed of light, while there is a parameter space 
for which the second mode is  subluminal: 
for instance, setting $q_1=3q_0/2$  results in a broad parameter range 
consistent with the "no ghosts" and  "no gradient instability" conditions,  
with one of the scalars strictly subluminal (the other one being  luminal).  
A representative part of the latter parameter space is shown in the right panel of Fig.\ref{fig:2}.

\paragraph{ \it \bf Discussion and outlook:}  The obtained solutions should have their counterparts  
in the full QMG theory,  beyond the decoupling limit. From our results it follows that the full QMG 
solutions should not have any ghost instabilities,  and {\it if} there is any tachyon instability, it can only be 
very slow,  with the time scale $m^{-1} \sim H_0^{-1}\sim 14\times 10^9$ years.  
The full QMG  solutions should  turn into those of pure massive gravity 
as $\omega$ is taken to infinity.   While the DL solutions discussed here 
are homogeneous and isotropic, they are such due to the diff invariance for $h_{\mn}$, and  
Galilean invariance for either  $\pi$ or $\sigma$. The Galilean invariance, being 
exact in the DL, is only an approximate symmetry in the full theory. Therefore,
the full theory solutions are likely to be  only  approximately
homogeneous and isotropic; this is similar to pure massive gravity where such 
metrics approximate well the standard homogeneous and isotropic 
evolution \cite {D'Amico:2011jj}, when the Vainshtein mechanism is at 
work\footnote{Moreover, QMG solutions for the open universe may even be homogeneous 
and isotropic, as in pure massive gravity \cite {Gumrukcuoglu:2011ew}.}.
Since our theory is that of  special bi-Galileons, the Vainshtein 
mechanism  is expected  to work for a large  domain of the parameter space. 
Moreover, the Higuchi problem need not arise as the St\"ueckelberg fields are necessarily 
inhomogeneous/anisotropic. We will report on this in  \cite {gkp}.

One may wonder whether quantum corrections can modify the classical picture given above - \emph{e.g.} alter the 
viable parameter space given in Fig. \ref{fig:2}.  The answer is likely to be 
negative, due to the non-renormalization properties of Galileons \cite{Luty:2003vm,*Hinterbichler:2010xn},  dRGT massive gravity \cite {deRham:2012ew,*deRham:2013qqa},  and related theories; these results can be summarized in the following statement: \emph{None of the couplings in the Lagrangian \eqref{eq:mgmasteraction} runs with energy below the scale $\Lambda_3$ in the decoupling limit.} Terms with at least two derivatives per field can certainly be generated via quantum loops, however they do not contribute to the equations of motion on the backgrounds considered here.  These non-renormalization properties guarantee that 
any possible tuning of the parameters $m,a_{2,3}$ and $\omega$, is  
technically natural \cite {deRham:2012ew,*deRham:2013qqa}. 

\paragraph{ \it \bf Acknowledgements:} We would like to thank Shinji Mukohyama and 
Enrico Trincherini for valuable communications.  
GG is supported by  NSF and the  NASA grant NNX12AF86G S06; 
R.K. is supported in part by a Grant-in-Aid for JSPS Fellows, and DP is supported in part by MIUR-FIRB grant RBFR12H1MW.

\bibliography{bibliography}

\end{document}